# Spin Fluctuations Yield zT Enhancement in Ferromagnets

Md Mobarak Hossain Polash,[1,2] Daryoosh Vashaee[1,2*]

[1]Department of Materials Science and Engineering, NC State University, Raleigh, NC 27606, US
[2]Department of Electrical and Computer Engineering, NC State University, Raleigh, NC 27606, US

## SUMMARY

Thermal fluctuation of local magnetization in magnetic metals intercoupled with charge carriers and phonons offers a path to enhance thermoelectric performance. The thermopower enhancement by spin fluctuations (SF) has been observed before. However, the crucial evidence for enhancing thermoelectric-figure-of-merit (zT) by SF has not been reported until now. Here we report evidence for such enhancement in the ferromagnetic CrTe. The SF leads to nearly 80% zT enhancement in ferromagnetic CrTe near and below $T_C$~335 K. The ferromagnetism in CrTe is originated from the collective electronic and localized magnetic moments. The field-dependent transport properties demonstrate the profound impact of SF on the electrons and phonons. The SF simultaneously enhances the thermopower and reduces the thermal conductivity. Under an external magnetic field, the enhancement in thermopower is suppressed, and the thermal conductivity is enhanced, evidencing the existence of a strong SF near and below $T_C$. The anomalous thermoelectric transport properties are analyzed based on theoretical models, and a good agreement with experimental data is found. Furthermore, the detailed analysis proves an insignificant impact from spin-wave contribution to the transport properties. This study contributes to the fundamental understanding of spin fluctuation for designing high-performance spin-driven thermoelectric materials.

**Subject Areas**

Spin fluctuation, transition-metal chalcogenide, magnetic thermoelectrics, spin-caloritronic, spin-driven thermoelectrics

## INTRODUCTION

Enhancing the entropy flow via electron gas by exploiting the intercoupled transport of electron, phonon, and spin has motivated a considerable amount of research recently in the thermoelectric society, which opened up a new direction in waste energy harvesting known as spin-caloritronics.[1,2,3] Spins, the fundamental entropy carriers on electronic orbitals as a quantum nature of electrons and lattice ions as collective spin excitations, a.k.a magnons, offer a degree of freedom to engineer the counter-indicative thermoelectric material properties, namely electrical conductivity (σ), thermal conductivity (κ), and thermopower (α) to design high-performance spin-driven thermoelectric materials.[4,5,6] The ever-growing research interests in thermoelectric-based green energy harvesting have forcefully emerged numerous pathways for designing materials for carbon-free energy harvesting applications with the broad societal needs of mitigating greenhouse and ozone-depletion potentials. Magnetic and paramagnetic semiconductors, among these pathways, offer a relatively new landscape with a large number of materials for investigations.[7,8,9] Therefore, designing high-performance thermoelectrics requires a deep understanding of the spin-driven effects. For this purpose, the physics learned from the simple elemental or binary magnetic materials can shed light on the underlying spin-driven transport natures without mixing the contributing effects due to a complex structure. The transition-metal chalcogenides (TMCs) are considered prospective exotic material families for versatile spin-based materials applications.[10,11,12] Consequently, they have been extensively studied as thermoelectric materials for their strong spin and quantum driven properties along with their salient electronic and phononic transport properties.[4,5,11,13,14] The interplay between electronic itinerancy and localized magnetization in metallic TMCs can drive different promising spin-caloritronic effects such as magnon-drag,[15,16] paramagnon-drag,[4,5] spin-entropy,[17,18] and spin-fluctuation.[19,20] Most of these spin-caloritronic effects provide enhancement primarily in the thermopower. Magnon-drag, for example, enhances the thermopower in the magnetically ordered domain and increases near the phase transition temperature.[21,22,23] However, the paramagnon-drag enhances the thermopower, and consequently the figure-of-merit ($zT = T\alpha^2\sigma/\kappa$, $T$ is the temperature), in the disordered magnetic domain where the short-range magnetic ordering helps to survive the spin-wave packets, a.k.a paramagnon quasiparticles, that act like magnons to itinerant carriers under certain conditions.[4]

---

* Email: dvashae@ncsu.edu

Despite the advances in thermoelectric materials over the last two decades, the zT is still low to compete with alternate technologies and must be improved to make the commercial thrive. Therefore, the daunting energy demand of the future sustainable society urges the deployment of highly efficient thermoelectric devices. Such enhancement can be achieved only by optimizing the thermoelectric properties from the synergistic engineering of all degrees of freedom.[24,25,26] The electron and phonon transport engineering methods have already been extensively explored. As a result, a significant zT improvement has been achieved; however, the progress has reached a plateau with little progress over the last several years. Therefore, other strategic degrees of freedom based on spin and quantum effects have taken attention recently to be explored as a new route for improving zT.[4,21,23,27,28]

With the abovementioned objectives, the impact of the spin-fluctuation effect on thermoelectric transport properties of CrTe, a binary TMC with simple NiAs ferromagnetic (FM) structure, is investigated to learn the physics of the underlying spin and quantum driven effect as a new strategy to enhance the zT. The prospect of spin-fluctuation induced thermopower enhancement and hence in thermoelectric power factor ($P = \alpha^2\sigma$) has been reported recently without exploring the overall impact of spin-fluctuation on zT.[23,29,30] We explore and weigh the impact of spin-fluctuation on all the parameters entering zT - electrical conductivity, thermal conductivity, and thermopower. We show that an external magnetic field suppresses the SF, leading to a significant reduction in the thermopower and increased thermal conductivity. Consequently, once an external magnetic field suppresses the SF, the zT reduces significantly. Overall, around 60-80% enhancement in zT has been observed in CrTe from the SF at zero-field near and below the transition temperature, which is remarkable compared to the previous reports.[31,32] This observation implies that the SF can be a likely source of zT enhancement in magnetic thermoelectric materials. Furthermore, the physics learned about the SF-mediated electrical and thermal transport properties from this work can be instrumental for the broader scientific community interested in spin-driven research for versatile application fields.

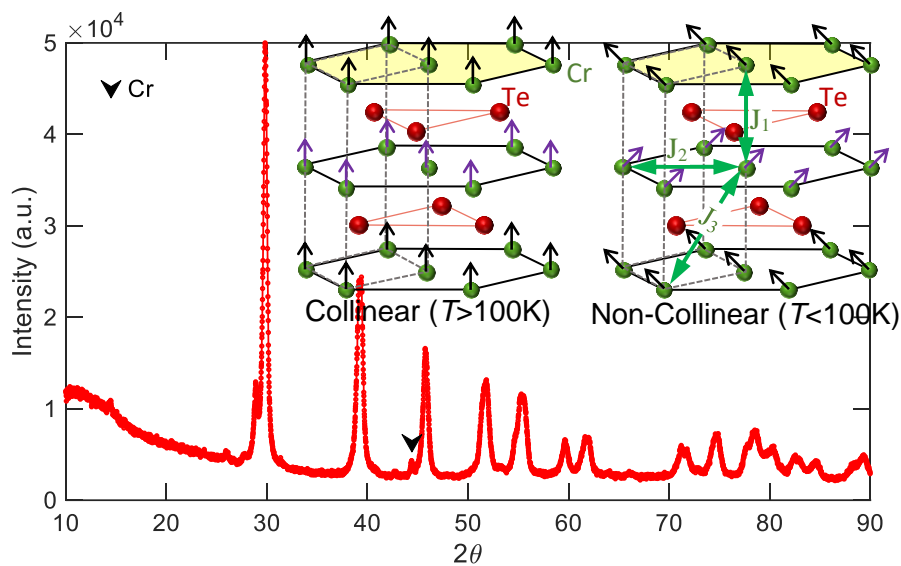

**Figure 1: Chemical Structure of CrTe from XRD**
The room temperature XRD pattern of CrTe. The inset shows the NiAs crystal structure along with their collinear and non-collinear spin orientation and exchange energies between nearest, next-nearest, and next-next-nearest Cr-Cr ions. Non-collinear spin orientation can exist below 100 K.

## RESULTS AND DISCUSSION

### Chemical Phase Analysis and Magnetic Properties

Phase identification of the synthesized CrTe is performed on the room temperature XRD data, shown in Figure 1, using the PDXL software. The phase analysis indicates the formation of CrTe with a small trace of Cr, which agrees with previous literature.[33] The presence of a small amount of Cr infers the formation of the $Cr_{1-\delta}Te$ phase, but for simplicity, we used the CrTe symbol for further discussion in the paper. The lattice parameters extracted from the XRD are $a,b = 3.97$Å, $c = 6.19$Å, $\alpha,\beta = 90°$, and $\gamma = 120°$, which have small temperature-dependency.[34] The reference intensity ratio (RIR) analysis calculates the presence of ~1% Cr in CrTe. CrTe is an FM hexagonal crystal (Curie temperature, $T_C \approx 340$ K) with NiAs structure and metallic conductivity.[35,36] CrTe has a distorted hexagonal close packing of Te atoms with Cr atoms in octahedral interstices. CrTe typically has Cr-vacancies in every second metal layer. The number of vacancies can introduce Te-rich CrTe phases ($Cr_{1-\delta}Te$) with a distortion in the crystal structure, affecting the material properties, including magnetic parameters.[34,35] The zinc-blend CrTe shows half-metallic FM nature that originated from the interplay between exchange splitting and hybridization.[37] In general, NiAs-type 3d chalcogenides have both direct exchange interactions between nearest-neighbor 3d ions and indirect exchange in the form of superexchange interaction between two 3d ions via an intermediated anion (shown in Figure 1).[33] The impurity phase Cr is antiferromagnetic (AFM) below ~310 K.[38] The net magnetic moment in AFM Cr is nearly zero with an insignificant impact on the magnetic properties of CrTe.

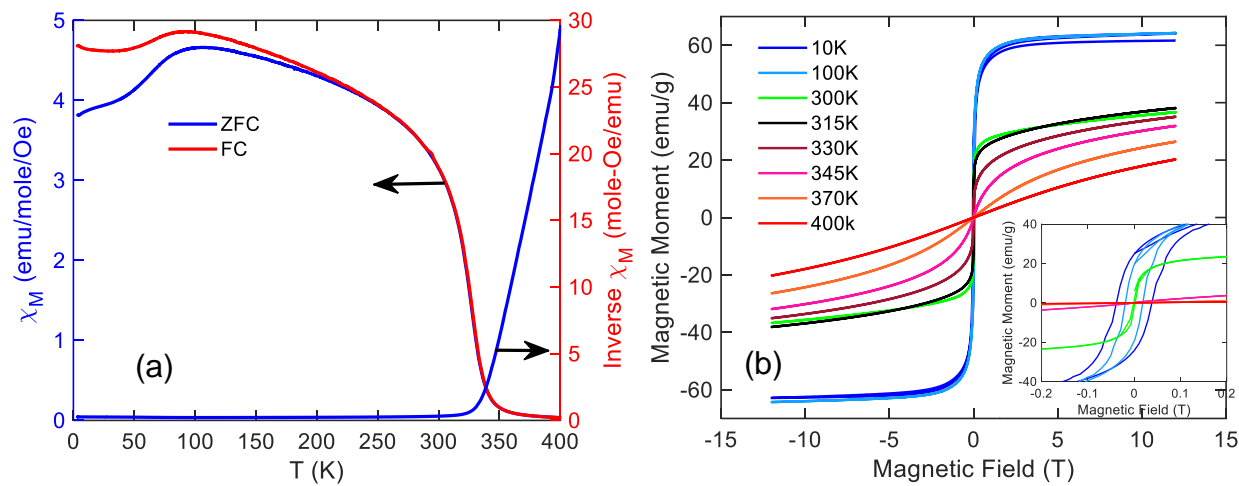

**Figure 2: Magnetic Properties of CrTe**
Illustration of (a) temperature-dependent zero-field cooled (ZFC) and field-cooled (FC) magnetic susceptibility ($\chi$) and inverse magnetic susceptibility, and (b) field-dependent magnetic susceptibility at different temperatures. Inset shows the close view of coercive field and remnant magnetization trends.

To investigate the magnetic nature of the CrTe sample, we measured both temperature and field-dependent magnetic susceptibility, as illustrated in Figure 2. The temperature-dependent magnetic moment of CrTe exhibits the nature of FM. Interestingly, both zero-field cooled (ZFC) and field-cooled (FC) molar magnetic susceptibility show similar trends above ~200 K, while FC susceptibility is larger below 200 K. At $T < 100$ K, magnetic moment decreases gradually also observed in previous work.[35,39] The decrease in the magnetic moment at a lower temperature can be attributed to the canted structure coming from the collinear to non-collinear structural transition shown in *Figure 1*.[39] Other possible reasons can be the itinerant FM origin[39] and the softening of the magnon modes due to the drop of the spin-wave stiffness.[40] The softening of the magnon mode can be associated with the canted structure. The 3d orbitals of Cr in CrTe contribute to the formation of localized and band carriers and, therefore, can originate the bulk and itinerant ferromagnetism in CrTe, respectively.[35,36,39]

The inverse magnetic susceptibility is calculated to extract several magnetic parameters using the Curie-Weiss law in the paramagnetic domain. The Curie-Weiss law reveals that the CrTe sample is ferromagnetic with $T_c \sim 336$ K, Curie constant, $C = 2.32$ emu-K/Oe/Mole, effective paramagnetic moment, $\mu_p = 4.3\ \mu_B$, and effective spin number, $S = 1.71$. The extracted parameters are within the range of previously reported values.[35,36] To understand the FM nature of the sample, saturation magnetization is calculated as $\mu_s \sim 2.13\ \mu_B$ at 10 K from field-dependent magnetic susceptibility. The ratio between effective paramagnetic and saturation magnetization ($\mu_p/\mu_s$) can indicate the FM nature of the materials. For CrTe, the $\mu_p/\mu_s \approx 2$ means that the FM nature of CrTe is originated from collective electron and localized magnetic moments.[41] Note that a unity ratio is the indicator of a purely localized moment-induced FM nature.[41] The intermediate ferromagnetism can be explained by itinerant ferromagnetism.[42] Moreover, the heat capacity data shows the spin-wave in the system - note that the spin-wave originates from the localized magnetization. Therefore, both itinerant and bulk ferromagnetism is considered the origin of the collective FM nature of CrTe, where 3d Cr ions can be the origin for both itinerant and bulk magnetism. The field-dependent magnetic susceptibility exhibits the characteristics hysteresis nature of FM phase below 336 K and no hysteresis in paramagnetic phase above ~336 K. Moreover, field-dependent magnetic susceptibility exhibits saturation with the field at 10 K and unsaturated moment near $T_C$, which can be associated with the itinerant ferromagnetism.

Here, it is essential to discuss the difference between localized magnetization-induced ferromagnetism and electronic itinerancy-induced nearly or weakly ferromagnetism in metals. The localized magnetization-induced FM phase below $T_C$ originates from the thermally fluctuated coupled localized moments known as spin-wave or magnons.[43] In the paramagnetic phase, the short-lived spin-wave excitations are called paramagnons which are overdamped due to the weakly coupled moments.[43] This type of FM material follow the Curie-Weiss law and has $\mu_p/\mu_s \approx 1$.[43] On the other hand, the nearly or weakly ferromagnetism is coming from the non-uniform distribution of itinerant electron-hole spin bands.[43] Nearly or weak FM metals exhibit Stoner type excitations with unsaturated moments with temperature and magnetic field, have $\mu_p/\mu_s > 1$.[43] In localized magnetic moment-induced FMs, the SF is caused by the spin-related scattering on thermally fluctuated propagating magnons, while in nearly FMs, the SF is occurred by exchange-enhanced non-propagating spin-density fluctuations.[44] Both types of sources for SF mechanisms can exist in transition-metal systems depending on spin-wave dispersion and Stoner excitation continuum nature for temperature and spin wavevector.[44,45]

**Magneto-thermoelectric transport Properties**

In this work, the magneto-thermoelectric properties of CrTe are measured to observe the impact of the spin-caloritronic effect in CrTe. Therefore, the measurements are performed in the range of 4-380K at 0 T, 4 T, 8 T, and 12 T fields, as illustrated in *Figure 3*. The transport properties are calculated from the averaged data of the positive and negative field-dependent measurement, which helps to eliminate the parasitic field-induced effects. The measurements of magneto-thermoelectric transport properties include thermopower, thermal conductivity, and electrical conductivity. zT is calculated from these properties. According to *Figure 3*(a), the thermopower at zero field shows almost a linear trend with a negative value in the FM domain. The slope changes at ~335K, which is also the magnetic transition temperature ($T_c$~336 K) of CrTe. Under the fields, the thermopower of CrTe reduced significantly by more than 20% at $T_c$. The field-dependent suppression in thermopower is about the same for all the applied fields. Since the FM nature of CrTe is due to the itinerant electrons and the localized magnetic moments, the spin-dependent scattering originated from the propagating spin-wave, and the non-propagating spin-density fluctuations can influence the thermoelectric properties. Besides the magnetic contributions, the carrier diffusion also contributes to the thermopower. The diffusion part can be extracted from the low-temperature trend of the thermopower at 12 T, as the high field suppresses the spin-based thermoelectric effects. A detailed analysis of the magnon-drag and SF is required to evaluate the contributions of different spin-driven effects on thermopower, which is discussed later.

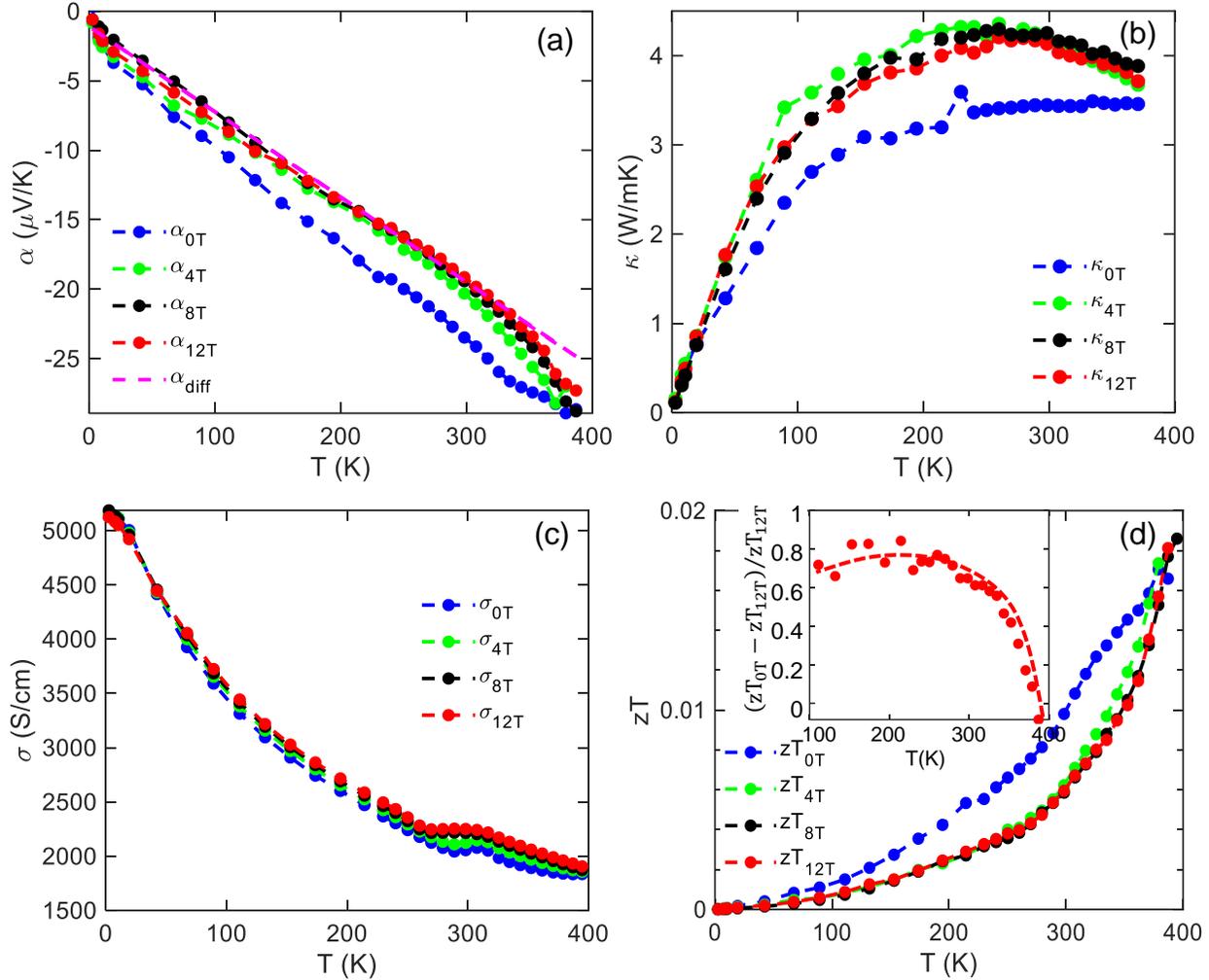

**Figure 3: Thermoelectric Transport Properties of CrTe**
Magneto-thermoelectric transport properties of CrTe at different fields i.e., 0T, 4T, 8T, and 12T: (a) thermopower (α), (b) thermal conductivity (κ), (c) electrical conductivity (σ), and (d) zT. The inset of (d) shows the field-induced suppression in zT which indicates the prospect of spin-fluctuations in spin-driven thermoelectrics.

Unlike the thermopower, the thermal conductivity shows field-induced enhancement in the FM domain, although it approaches the zero-field thermal conductivity trend at higher temperatures. Characteristically, the spin-based contribution in thermal conductivity of magnetic or paramagnetic materials can be both positive or negative. The magnon quasiparticles positively contribute to heat conduction added to the electronic and lattice heat conductions in magnetic materials. At a sufficiently high field, the FM magnons can be suppressed by damping out the thermal excitations of localized magnetic moments, which leads to the suppression of the magnonic thermal conductivity with the field. On the other hand, the presence of SF in magnetic or paramagnetic systems typically introduces spin-phonon scattering, which reduces the lattice thermal conductivity. The SF can also reduce the electronic thermal conductivity due to the presence of spin-related scattering. With the field-induced suppression of SF, the overall thermal conductivity enhances, explaining the field-dependent thermal conductivity enhancement in CrTe.

*Figure 3*(c) illustrates the field-dependent electrical conductivity. It can be seen that the electrical conductivity increases with the field. In the presence of both magnons and SF, the conduction carriers can experience excess scattering due to the existence of spin-flip and spin non-flip scatterings. The spin-disorder scattering enhances with temperature and maximizes at the transition temperature. The spin-disorder scattering directly relates to the carrier mobility, and hence, the electrical conductivity dominates by the trend of spin-disorder scattering-mediated carrier mobility. With the field, the suppression of the spin-disorder scatterings causes enhancing the electrical conductivity. The spin-driven effects on electrical conductivity are further discussed with the galvanomagnetic properties, including Hall and magnetoresistance later. Overall, spin-driven effects in CrTe yield a significant enhancement in the zT at zero-field, which is suppressed with the field (see *Figure 3*(d)). A significant enhancement is observed in the temperature range of 100-375 K. The nearly zero difference in zT at T > 375 K infers that the thermal fluctuations of localized magnetic moments at a higher temperature overcome the field-induced quenching. According to the inset of *Figure 3*(d), the spin-driven effects can enhance the zT by 60-80% near and below $T_c$. As discussed, the magneto-thermoelectric properties can be influenced by both the magnon/paramagnon-drag and SF. Therefore, it is essential to evaluate the contributions of each effect. The field-dependent heat capacity analysis can shed light on the relative contributions of spin-wave and SF. It contains the features from different entropy carriers in the system, discussed in the next section.

**Field-dependent Heat Capacity Analysis**

Heat capacity is one of the reliable and direct methods to probe the existence of different entropy carriers, including electronic, phonons, magnons, Schottky, dilation, spin-state transitions, and spin fluctuations.[46,47] All of the contributing sources in heat capacity have distinct temperature dependency.[46,47] Among these sources, magnetic heat capacity components can show strong magnetic field dependency based on the magnetic nature, provided that they can be quenched or modified with the external magnetic fields.[46,47] Quenching of magnonic heat capacity depends on the spin-wave stiffness, exchange energy, spin number, and magnetization.[6] On the other hand, spin fluctuation contribution can be quenched by suppressing the inelastic spin-flip scattering.[47] If the external field is sufficiently large to create the Zeeman energy splitting comparable or larger than the characteristic spin fluctuation energy, the available thermal energy cannot introduce SF in the system.[47] As mentioned in the earlier section, the metallic FM CrTe has collective contributions from itinerant and localized electrons. Therefore, it has magnetic contributions to the heat capacity from both magnons and SF. They can be extracted by modeling the nonmagnetic heat capacity components and subtracting them from the total heat capacity.

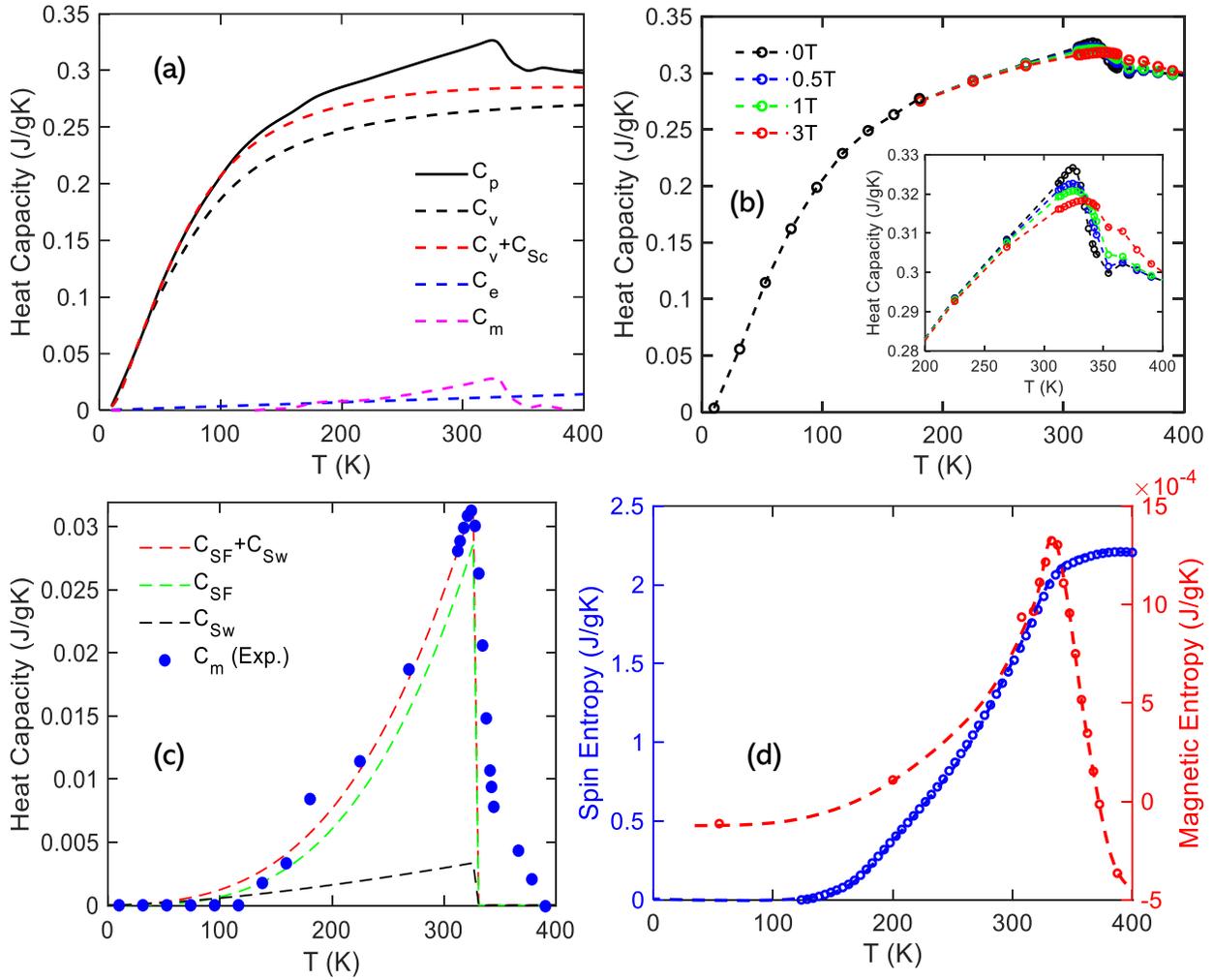

**Figure 4: Heat Capacity and Entropy properties of CrTe**
Illustration of the heat capacity of CrTe: (a) different heat capacity contributions at zero field obtained theoretically and experimentally, (b) field-dependent heat capacity measured at 0 T, 0.5 T, 1 T, and 3 T, (c) theoretically and experimentally extracted different magnetic heat capacity contributions from magnons (spin-wave) and spin fluctuations, and (d) spin entropy obtained from the magnetic heat capacity and magnetic entropy calculated from field-dependent magnetic susceptibility shown in **Figure 2**(b).

Figure 4(a) presents the zero-field heat capacity ($C_p$) and the contributions from phonon ($C_v$), electron ($C_e$), Schottky ($C_{Sc}$), and magnetic ($C_m$). The theoretical models associated with respective heat capacity contributions are given in the supplementary of previous literature.[10] The physical and fitting parameters associated with heat capacity modeling are listed in the table (see STAR Methods). The insignificant variation in the thermal expansion results in an insignificant heat capacity contribution from dilation.[34] The magnetic heat capacity illustrated in Figure 4(c) shows a peak at ~332 K, which is also near $T_C$~336 K. To extract the individual magnetic contributions, the magnonic heat capacity originated from the spin-wave and SF is estimated from:[47,48,49]

$$C_m = C_{SW} + C_{SF} = cNk_B \left(\frac{k_B T}{2JS}\right)^{1.5} + DT^3 \ln T \qquad (1)$$

Here, $C_{SW}$ and $C_{SF}$ are the spin-wave and SF contributions to magnetic heat capacity ($C_m$), $c$ is the constant related to the crystal structure, $J$ is the exchange energy, $N$ is the Avogadro's number, $k_B$ is the Boltzmann constant, $S$ is the ground spin number, and $D$ is a material-dependent constant. According to Figure 4(c), the SF contribution provides an anomalous enhancement in the electronic contribution to the heat capacity, which is more significant than the spin-wave contribution. External fields can significantly suppress the heat capacity due to SF. The electronic heat capacity enhancement due to SF is typically due to the renormalization of the effective electron mass and the variation in electron self-energy with temperature.[50]

Figure 4(b) illustrates the field-dependent heat capacity trends. It is observed that the magnetic heat capacity features are significantly suppressed with a field. The suppression of magnetic heat capacity can be associated with the suppression of both magnons and SF. The field-induced Zeeman splitting can also modify the Schottky heat capacity contribution. As the thermopower is related to the thermodynamic entropy of the system carried by conduction carriers, the contributing entropy source can be estimated by calculating the available entropy from different sources. In CrTe, two types of spin-related entropy can exist: spin entropy from the spin-wave and magnetic entropy from the magneto-caloric effect. The insertion and removal of the external field can introduce a magneto-caloric effect which maximizes at the transition temperature. Therefore, we estimated both entropy contributions from associated relations and illustrated them in Figure 4(d).

The spin entropy ($\Delta S_{SW}$) is estimated from the heat capacity from: $\Delta S_{SW} = \int \frac{C_{SW}}{T} dT$, and the magnetic entropy ($\Delta S_M$) is calculated from field-dependent magnetic susceptibility using the relation: $\Delta S_M = \int \left(\frac{\partial M}{\partial T}\right)_H dH$, where $H$ is the magnetic field. It is seen in Figure 4(d) that the magnetic entropy is orders of magnitude lower than the spin entropy. Moreover, the magnetic entropy decays quickly below and above the transition temperature. Thus, the trend in spin entropy rather closely follows the trend of thermopower, which indicates that the spin-related contribution in thermopower is coming from magnons and spin fluctuation.

The overall conclusions from the field-dependent heat capacity analysis are: (i) both spin-wave and spin fluctuation are present in CrTe with the spin fluctuation being dominant, (ii) the magneto-caloric effect has an insignificant impact on the thermoelectric transport of CrTe, and (iii) the field-induced modification in heat capacity can be due to the quenching of the spin-wave and spin fluctuation. Besides heat capacity, galvanomagnetic properties can also shed light on the spin and quantum nature of conduction carriers discussed in the following sections.

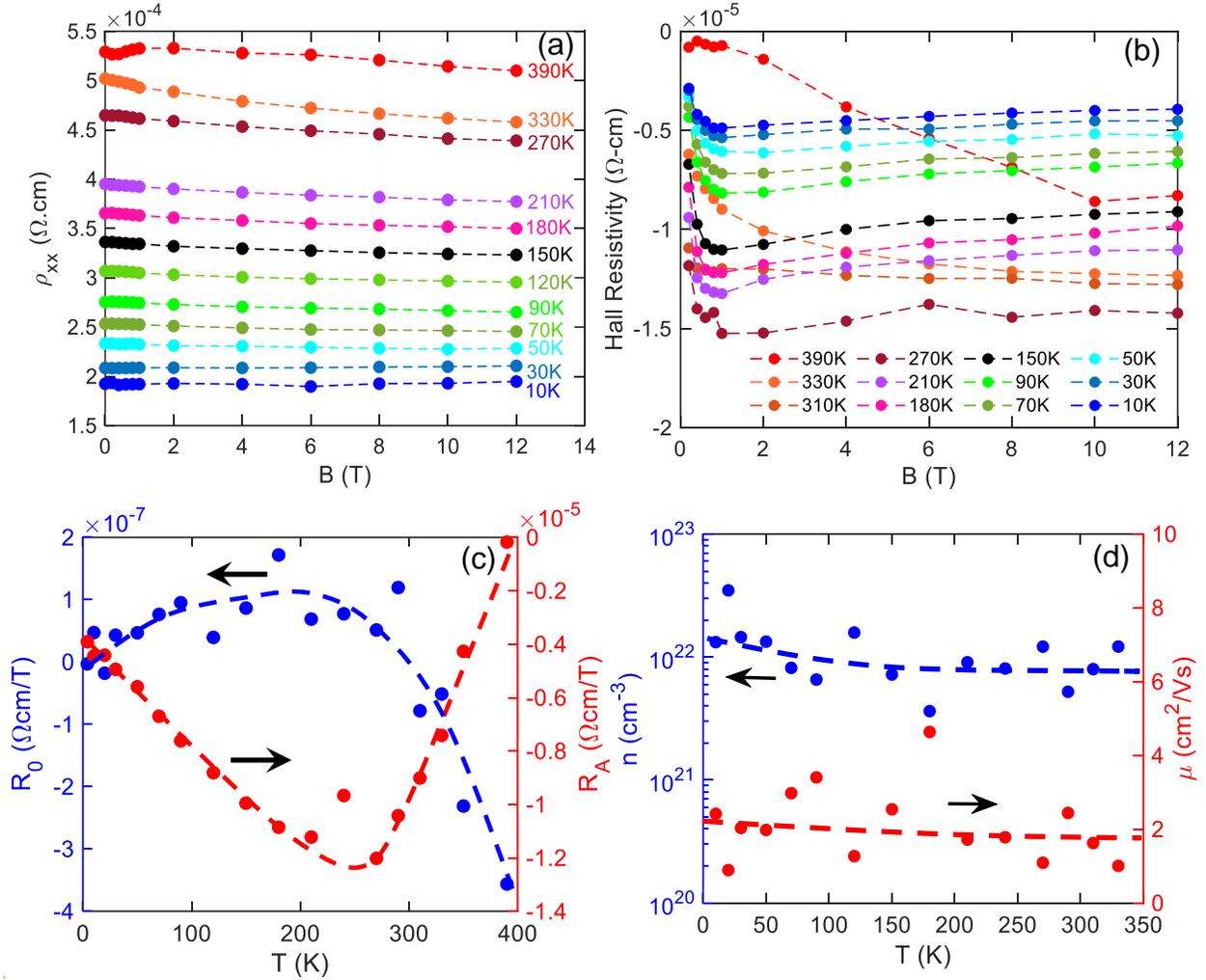

**Figure 5: Galvanomagnetic Properties of CrTe**
Illustration of Hall Properties of CrTe: (a) field-dependent electrical resistivity at different temperatures, (b) field-dependent Hall resistivity at selected temperatures, (c) the ordinary and anomalous hall coefficient, and (d) Hall carrier concentration and mobility.

## Galvanomagnetic Properties

The isothermal galvanomagnetic properties, including the Hall effect and transverse magnetoresistance (TMR), reveal the carrier transport nature. The Hall properties provide information on the type and concentration of charge carriers. The field-dependent variation of the resistance yields information on the interaction between carriers and other transport carriers and the Fermi surface.[51] The isothermal condition eliminates the contribution from adiabatic galvanomagnetic effects due to the heat current-induced galvanomagnetic potentials.[52] The isothermal galvanomagnetic properties of the sample are measured with the Van der Pauw (VdP) method over the 10 K-400 K temperature range and -12 T to 12 T magnetic field in Physical Property Measurement System (PPMS) made by Quantum Design. The isothermal Hall and TMR properties are estimated from the average of the positive and negative field data, as illustrated in Figure 5 and Figure 6.

The field-dependent longitudinal and transverse resistivity (Hall resistivity) shown in Figure 5(a)-(b) reveals the existence of both ordinary and anomalous Hall effects, which is common in FM materials. The ordinary and anomalous Hall coefficient (OHC and AHC) are estimated semi-empirically from, $\rho_{xy} = R_H = R_O B + R_A \mu_0 M$, where $\rho_{xy}$ or $R_H$ is the transverse resistivity or Hall resistivity, $R_O$ and $R_A$ are OHC and AHC, respectively, $\mu_0$ is vacuum permeability, $B$ is the magnetic field, and $M$ is the magnetization. While the ordinary Hall effect is due to the transverse Lorentz force under the external magnetic field, the anomalous Hall effect is caused by the breaking of the time-reversal symmetry due to the intrinsic or defect-induced spin-orbit coupling forces.[53] In CrTe, the anomalous Hall effect is more dominant than the ordinary Hall effect due to the strong interaction between conduction and localized electrons.[54] The OHC is positive in the FM domain due to the hole conduction[54] and shows a sign change near $T_C$. The hole conduction nature in CrTe has already been reported in previous literature.[35,36,54,55] Therefore, the sign change can be attributed to the temperature-independent spin-orbit interaction between conduction and magnetic electrons.[54] On the other hand, AHC is always negative, has a broad peak at ~270 K, and remains non-zero above the Curie temperature. This indicates the presence of a spin-orbit coupling field below and above the Curie temperature. The carrier concentration and mobility are calculated from OHC and electrical conductivity considering the single-band Hall transport. The Hall carrier concentration in CrTe is obtained in the range of $1 \times 10^{22}$ cm$^{-3}$ which is close to the theoretically estimated concentration of 0.2 holes/per unit formula.[35] The variation can be associated with the variation in vacancies and impurity phases present in CrTe. The carrier mobility is found as ~2 cm$^2$/Vs.

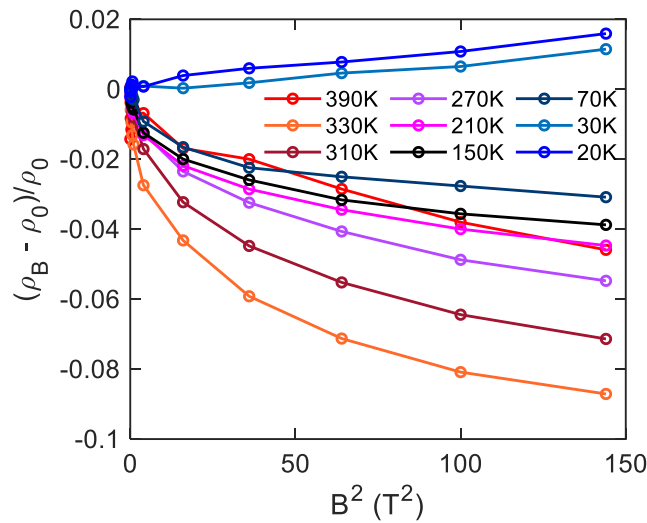

**Figure 6: Transverse Magnetoresistance of CrTe**
Transverse magnetoresistance (TMR) as a function of square of external magnetic fields at different temperatures below and above the transition temperature.

Figure 6 illustrates the TMR (current flow is perpendicular to the applied field) as a function of the square of the external magnetic field ($B^2$). In magnetic metals, both longitudinal and transverse magnetoresistance are significantly affected by the SF effect. Along with the SF contribution, the resistivity of FM metals can also have a contribution from domain structure, i.e., $\rho(T,B) \approx \rho_{domain}(T,B) + \rho_{SF}(T,B)$.[56] For bulk samples with large grain sizes, the positive change of magnetoresistance mainly comes from the domain contribution.[56] This increase of magnetoresistance at low fields (lower than the local anisotropic field, $H_A$) is generally due to the domain movement controlled by the $H_A$.[56] As the CrTe does not show any field-dependent enhancement in TMR at low fields, the domain contribution in TMR must be insignificant. However, according to Figure 6, the TMR below ~50 K shows a positive quadratic increase with the field (~$B^2$ dependency), while the TMRs above ~50 K always show negative trends. The positive quadratic nature of TMR below ~50 K can be associated with the dominance of the cyclotron motion of carriers along the open orbits on the Fermi surface, a.k.a. the trajectory effect. Above ~50 K, CrTe exhibits a typical TMR nature of an FM metal which shows a negative TMR originated from the magnetic field suppression of the spin fluctuations in space and time domains that leads to the reduction of spin-related scattering, decreasing the electrical resistivity under the field.[44,57]

Both heat capacity and TMR measurements suggest strong SF in CrTe, which directly impacts the carrier transport properties. Therefore, we discuss the SF and its impact on the thermoelectric properties of CrTe in the following section.

**Impact of Spin Fluctuations on Transport Properties**

As mentioned earlier, the spin-wave (or magnon) and exchange-enhanced spin fluctuation (or spin-density fluctuation) contributions can coexist in ferromagnetic metals and make distinct impacts on the transport properties. In such systems, the temperature and field-dependent magnetization is dominated by the propagating spin-wave at low temperatures and the non-propagating spin fluctuations at intermediate temperatures and temperatures near the Curie point.[44] The impacts of magnons and SF on transport properties are determined by the spin-dependent scattering arising from the charge-spin interaction[5] and the correlated electron-hole pair collective excitations.[44] The magnetic excitations in an FM metal with the collective origin of ferromagnetisms is both temperature and energy dispersion-dependent. At zero temperature, magnetic excitations due to spin-wave are confined around zero magnon wavevector (q=0) in the Brillouin zone, representing the bound states for electron-hole pairs.[44] The energy gap between low-energy spin-wave and bound states corresponds to the energy splitting between the majority and minority spin bands (shown in Figure 7).[44] This energy gap reduces with the increase in $q$, and at certain $q=q_C$, the magnon dispersion curve enters the Stoner excitation continuum. At $q>q_C$, magnons get damped. The dominant magnetic excitation mechanism is transitioned from spin-waves to exchange-enhanced spin fluctuations, observed in thermodynamic quantities like magnetization, electrical conductivity, thermal conductivity, heat capacity, and thermopower at specific temperature ranges.[44] Therefore, at low temperatures ($T<<T_C$), the dominant magnetic contribution in transport properties is due to the long-wavelength low-frequency spin-wave modes, while at intermediate temperature ($T<T_C$) and near $T\sim T_C$, SF has a dominant contribution in the transport properties.[44]

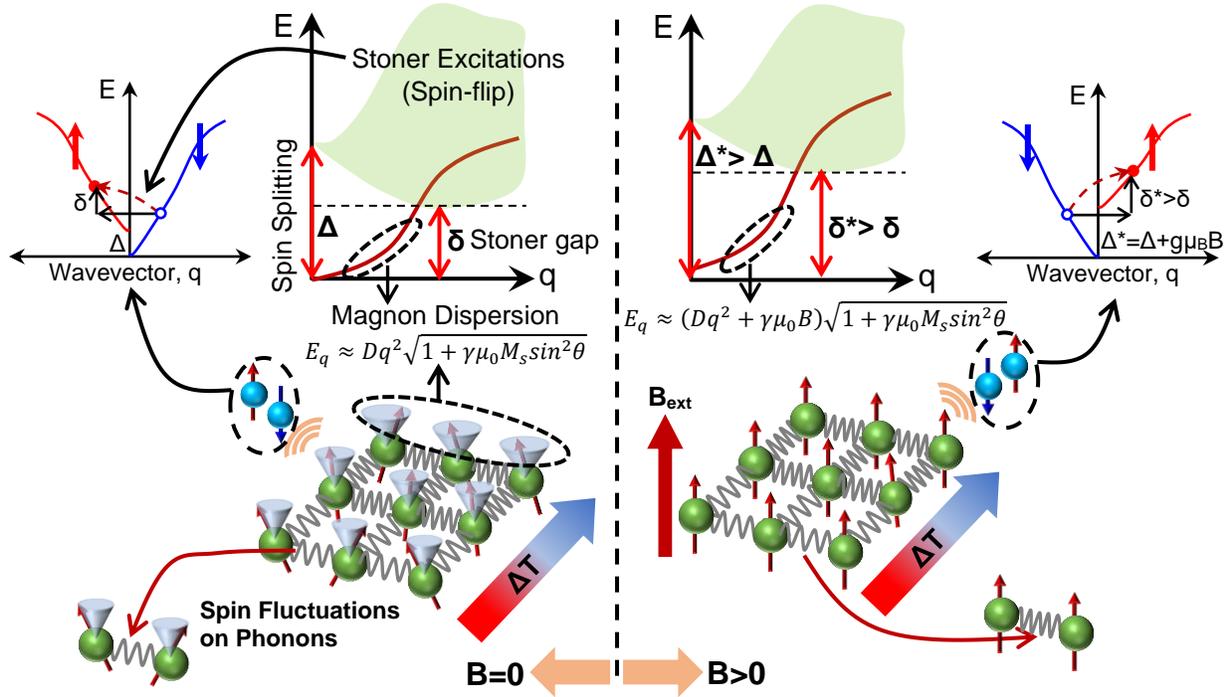

**Figure 7: Fundamentals on Spin Fluctuations in absence or presence of external magnetic field**
Illustration of spin fluctuation (SF) and the spin-wave in the absence and presence of magnetic fields. SF exists in an itinerant FM system with a non-zero stoner gap between spin majority and minority bands. Temperature-dependent spin-flip scatterings or stoner excitations are the origins of the SF-influenced carrier transport. External field-induced Zeeman splitting can quench the Stoner excitations to certain temperatures. On the other hand, spin-wave is coming from the temperature-induced fluctuation of magnetization of lattice ions. Spin-wave excitations can drag the itinerant carriers by the momentum relaxation process. Spin-phonon scattering can also have a direct impact on phonon transport properties.

Under external field, both spin-wave and SF get quenched, and hence, the magnetic excitations due to the spin-wave and SF are also suppressed. The field-dependent magnon dispersion can be expressed as:[58]

$$E_q \approx (Dq^2 + \gamma\mu_0 B)\sqrt{1 + \gamma\mu_0 M_s sin^2\theta} \qquad (2)$$

Where $D$ is the spin-wave stiffness, $\gamma$ is the gyromagnetic ratio, $M_s$ is the saturation magnetization, and $\theta$ is the angle between magnon flow and the external magnetic field direction. Under field, the magnon dispersion curve shifts towards higher energy with modified D* shown in Figure 7, increasing the energy gap for magnetic excitation or $q_c$.[44] Similarly, under field, the Stoner continuum shifts towards higher energy due to the increase in the spin-band splitting. Hence, higher thermal energy is required for the magnetic excitations (see Figure 7). Overall, both spin-wave and SF suppression can suppress the spin-dependent scattering on electrons and phonons to a specific temperature observed in the transport properties.

The field-dependent carrier transport nature in CrTe can shed light on the dominant magnetic excitation mechanisms. According to Figure 4(b) inset, the field-dependent heat capacity shows the suppression of magnetic contributions above ~225 K, while both spin-wave and SF exist above ~125 K with a significant contribution from SF. Therefore, the field-dependent suppression in magnetic heat capacity near $T_C$ is expected to be dominated by the SF quenching. The TMR is another property that manifests a distinct field- and temperature-dependent behavior. Generally, at low temperatures ($T<<T_C$), TMR due to the spin-wave exhibits $-Bln(B)$ dependency at the low fields and saturates at high fields.[44] With the temperature increase, the TMR due to spin-wave decreases quickly.[44] On the other hand, TMR due to SF at intermediate temperature and near $T_C$ exhibits ~$B^2$ dependency and saturates at a much higher field than the spin-wave mediated TMR.[44] Interestingly, TMR due to SF increases up to ~$T_C$.[44]

For the case of CrTe, according to Figure 6, TMR shows ~$B^2$ trends above 150 K, and the absolute value of TMR increases up to around $T_C$ and decreases beyond $T_C$. The TMR trend below 50K is due to the trajectory effect.[51] The trends support the presence of SF as a dominant magnetic excitation mechanism at intermediate and near Curie point temperatures.

The thermal transport properties of CrTe, including thermopower and thermal conductivity, also demonstrate distinct field-dependent features, as shown in Figure 3. Under the 12 T field, the magnetic excitations on carriers below $T_C$ are expected to be quenched completely. Hence, it can be considered that the thermopower at 12 T below $T_C$ is dominantly due to the diffusion thermopower. With this consideration, the magnetic contribution to the thermopower is extracted by subtracting the zero-field thermopower and thermopower at 12 T (shown in Figure 8).

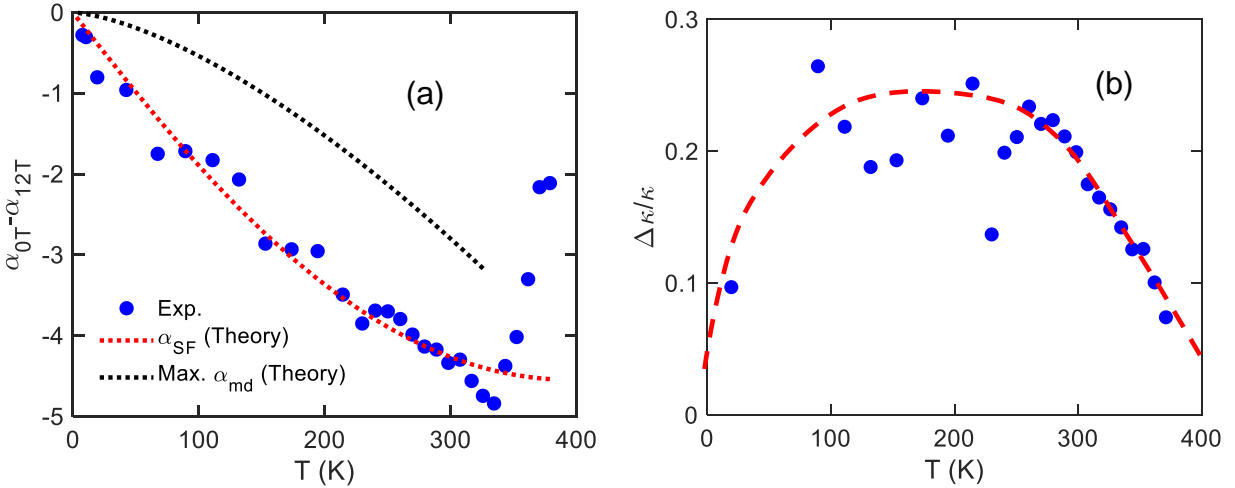

**Figure 8: Analysis of SF mediated thermopower and thermal conductivity**
(a) The magnetic contribution in thermopower of CrTe along with the theoretical contributions from spin-wave and spin fluctuations. Spin fluctuation shows better fit to the experimental data, and (b) the enhancement in thermal conductivity, $\Delta\kappa/\kappa = (\kappa(B=12) - \kappa(B=0))/\kappa(B=0)$, comparing to thermal conductivity at zero field, $\kappa(B=0)$.

The drag thermopower contribution due to FM magnons exhibits characteristic $T^{3/2}$ dependency and can be determined by magnonic heat capacity.[4] The magnon-drag thermopower ($\alpha_{md}$) is estimated from:[4]

$$\alpha_{md} = \frac{2}{3}\frac{C_{SW}}{ne}\frac{1}{1+\tau_{em}/\tau_m} \quad (3)$$

Where $n$ is the carrier concentration, $\tau_{em}$ is the magnon-electron relaxation time, and $\tau_m$ is the magnon lifetime. Considering the limit of $\tau_{em}/\tau_m$,[4] the maximum magnonic thermopower is achieved when $\tau_{em}/\tau_m = 2$ which is shown in Figure 8. According to the figure, the temperature-dependent trend and the maximum value of magnon-drag thermopower are characteristically different than the experimental trend. On the other hand, the experimental magnetic thermopower is fitted better with the SF-drag thermopower model, i.e.:[59]

$$\alpha_{SF}(T) = AT + BT\left(\frac{T}{T_0}\right)^2 \log\frac{\delta + \left(T/T_0\right)^2}{\left(T/T_0\right)^2} \quad (4)$$

Here $A$, $B$, $T_0$, and $\delta$ are the parameters obtained from fitting the experimental data with eq. (4). Considering the similar underlying physics of SF contributed heat capacity,[50] the first term in SF-drag thermopower is most likely coming from the SF-enhanced electronic heat capacity due to the SF-induced renormalization of electron effective masses, while the second term is associated with the variation in electron self-energy with respect to temperature. Here, the estimated parameters are $A = -0.02~\mu V/K^2$, $B = 0.0125~\mu V/K^3$, $T_0 = 815$ K, and $\delta = 4$. According to Figure 8, the theoretical SF-drag thermopower below $T_C$ closely follows experimental magnetic thermopower. The characteristic temperature of SF, $T_0$, is the temperature below which the SF contribution is dominant. $T_0$ can be close to the SF temperature, $T_{SF}$, often used in the literature.[59]

Like thermopower, the thermal conductivity ($\kappa$) manifests the existence of spin fluctuations that impedes the heat flow by electrons and phonons. Typically, thermal conductivity due to the magnons ($\kappa_m$) positively contributes to the total thermal conductivity that can be suppressed under an external field. On the other hand, the spin fluctuations reduce the thermal conductivity, so its suppression by the field can explain the thermal conductivity enhancement observed in Figure 8. This reduction can happen for both the electronic thermal conductivity ($\kappa_e$) and phononic thermal conductivity ($\kappa_{ph}$). Overall, in an FM metal with a collective origin of ferromagnetism, total thermal conductivity can have a contribution from charge carriers, phonons, SF, and magnons which can be expressed as:[60,61]

$$\kappa = \kappa_e + \kappa_{ph} + \kappa_m = \left(\frac{1}{\kappa_{e0}} + \frac{1}{\kappa_{e,SF}}\right)^{-1} + \left(\frac{1}{\kappa_{ph0}} + \frac{1}{\kappa_{ph,SF}}\right)^{-1} + \kappa_m \quad (5)$$

Here, $\kappa_{e0}$ and $\kappa_{ph0}$ are the electronic and phonon thermal conductivity without SF effect, and, $\kappa_{e,SF}$ and $\kappa_{ph,SF}$ are similar quantities but with SF effect. Due to the SF induced spin-phonon scattering, the total phonon mean free path is modified to $1/l_{ph} = 1/l_{mag} + 1/l_{nonmag}$, where the nonmagnetic contribution to mean free path ($l_{nonmag}$) is determined by the Umklapp phonon-phonon, electron-phonon, and phonon-defect scattering, and the magnetic contribution to the mean free path ($l_{mag}$) is limited by the spin-phonon scatterings due to SF.[60] Under an external magnetic field at finite temperature, suppression of SF leads to a larger $l_{mag}$ and hence, a larger $l_{ph}$ which provides higher thermal conductivity.[60] The SF can also impact the electronic thermal conductivity due to the SF-induced spin scattering on carriers. $\kappa_{e0}$ and $\kappa_{e,SF}$ show ~$T$ and ~$T^{-1}$ dependency, respectively.[61] The parallel thermal transport channels arising from SF effects on carriers and phonons cause the reduction in total thermal conductivity, while suppressing those transport channels under field enhances thermal conductivity. The suppression of the magnons, the third component contributing to the thermal conductivity, tends to reduce the thermal conductivity at high fields.

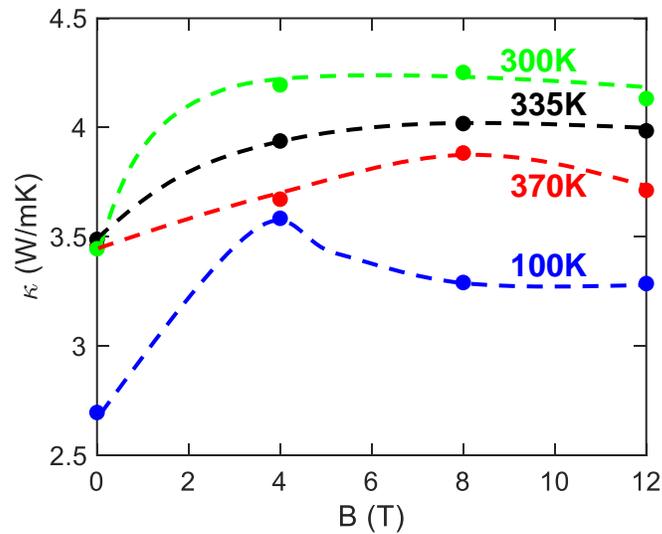

**Figure 9: Field-dependent thermal conductivity of CrTe**
Illustration of magnetic field-dependent thermal conductivity at different temperatures below and above the transition temperature.

According to Figure 9, the field-dependent thermal conductivity trends indicate both SF and magnon contributed thermal conductivity modifications. Typically, the SF-mediated phonon thermal conductivity trends follow the trend of the deviation of magnetization from the saturation value.[60] Comparing the magnetization trends shown in Figure 2, both electronic and phonon thermal conductivities must be affected by the SF effect. In contrast, the high field reduction in thermal conductivity can be associated with the field-induced reduction of the magnon thermal conductivity. Due to the complicated relationships of the magnetic contribution to thermal conductivity, the quantitative determination of each component is complex, and it requires a more detailed investigation beyond the scope of this study.

**Conclusion**

The interplay between electronic itinerancy and localized magnetization in transition-metal chalcogenide ferromagnets can introduce rich spin-dependent thermoelectric properties. The temperature, field, and energy-dependent spin-wave and spin fluctuation excitations manifest in the transport properties with characteristic features that help to understand the underlying mechanism of spin, carrier, and phonon transport. These intercoupled transport mechanisms recently exhibit several prospective spin-caloritronic effects, including paramagnon-drag, spin entropy, and spin fluctuation. This work demonstrates a significant zT enhancement in FM metal CrTe originated from the spin fluctuation mediated thermoelectric transport properties. The electronic, magnetic, heat capacity, galvanomagnetic, and thermoelectric transport properties are measured and analyzed with SF and spin-wave models. The experimental and theoretical investigations exhibit good agreement, indicating that the SF, and in this case more significantly than the spin-wave, can effectively enhance the zT in magnetic materials.

**Limitations of the study**

We did not discuss on how to engineer and enhance the SF mediated zT. This requires further studies to guide identifying or designing the material systems with a large zT resulted from SF. Moreover, a detailed study on calculating the individual contribution of the spin-wave, SF, and Schottky on transport and thermodynamic properties is suggested to provide a better understanding of the overall spin contributions in thermoelectric performance.

**ACKNOWLEDGMENTS**

This study is partially based upon work supported by the Air Force Office of Scientific Research (AFOSR) under contract number FA9550-19-1-0363 and the National Science Foundation (NSF) under grant numbers ECCS-1711253 and CBET-2110603.

**AUTHOR CONTRIBUTIONS**

D.V. directed the research. M.H.P. synthesized the samples and characterized the transport properties. M.H.P. performed the theoretical calculations and analyzed the results. All authors discussed the experimental results and contributed to prepare the manuscript.

**DECLARATION OF INTERESTS**

The authors declare no competing interests.